\documentstyle[11pt,newpasp,twoside,epsf]{article}
\def\edcomment#1{\iffalse\marginpar{\raggedright\sl#1\/}\else\relax\fi}
\reversemarginpar

\begin{document}
\title{Gamma-Ray Bursts as a Probe of the Epoch of Reionization}
\author{D. Q. Lamb$^1$ \& Z. Haiman$^2$}
\affil{$^1$University of Chicago, $^2$Princeton University}

\begin{abstract}

GRBs are expected to occur at redshifts far higher than the highest
quasar redshifts so far detected.   And unlike quasars, GRB afterglows
may provide ``clean'' probes of the epoch of reionization because no
complications (such as the presence of a strong Ly$\alpha$ emission
line) or ``proximity effects'' (such as the Str\"omgren sphere produced
by ionizing photons from the quasar) are expected.  Thus NIR and
optical observations of GRB afterglows may provide unique information
about the epoch of reionization.  In particular, the flux at
wavelengths shortward of Ly$\alpha$ provides a direct measure of the
density fluctuations of the IGM at the GRB redshift, while the flux at
wavelengths longward of Ly$\alpha$ provides an integrated measure of
the number of ionizing photons produced by stars in the host galaxy of
the GRB up until the burst occurs.  A comparison of the sizes of the
Str\"omgren spheres produced by stars in the host galaxies of GRBs and
by quasars then provides an estimate of the relative contributions of
star formation and quasars to reionization.  We use detailed
calculations of the expected shape of the GRB afterglow spectrum in the
vicinity of Ly$\alpha$ for GRBs at a variety of redshifts to illustrate
these points.

\end{abstract}

\vspace{-0.3in}
\section{Introduction}

Exactly when and how the universe was reionized are two of the most
important outstanding questions in astrophysical cosmology.  The
answers to these questions are of fundamental importance for
understanding the moment of first light, the formation of the first
galaxies, and the nature of the first generation of stars and of
quasars,

Spectral observations of Ly$\alpha$-emitting galaxies can be used as a
probe of the epoch of reionization (Haiman 2002).  However, such
galaxies are faint and exceedingly rare (Hu et al. 2002), and the
ability to draw inferences from the study of such Ly$\alpha$-emitting
galaxies is complicated by the necessity of disentangling the shape of
the trough due to the red damping wing of the Ly$\alpha$ resonance from
the (unknown) intrinsic profile of the Ly$\alpha$ emission line and the
continuum spectrum at nearby wavelengths of the galaxy (Haiman 2002).
In addition, scattering of the Ly$\alpha$ photons by a neutral IGM can
broaden the line, reducing its visibility and further complicating the
task of recovering the shape of the red wing of the Gunn-Peterson
trough.

Observations of bright quasars at high redshifts can also be used as a
probe of the epoch of reionization.  As an example, the recently
discovered bright quasar SDSS 1030+0524, which lies at $z = 6.28$,
shows a distinct Gunn-Peterson trough (Becker et al. 2001).  The lack
of any detectable flux shortward of $(1+z)\lambda_\alpha = 8850$ \AA\
implies a strong lower limit ($x_H > \sim 0.01$) on the mean
mass-weighted neutral fraction of the IGM at $z \approx 6$ (Fan et al.
2002).  This suggests that the IGM is neutral beyond $z \sim 7$.

However, such quasars are rare ($\sim 10^{-3}$ deg$^{-2}$) and
therefore difficult to find.  And again, the ability to draw inferences
from the study of such quasars is complicated by the necessity of
disentangling the shape of the trough due to the red damping wing of
the Ly$\alpha$ resonance from the (unknown) intrinsic profile of the
bright Ly$\alpha$ emission line (Madau \& Rees 2001; Cen \& Haiman
2001).

Figure 1 places GRBs in a cosmological context.  GRBs have several
distinct advantages over Ly$\alpha$-emission galaxies and quasars as a
probe of the epoch of reionization:

\begin{itemize}

\item
GRBs are by far the most luminous events in the universe, and are
therefore easy to find.

\item
If GRBs are produced by the collapse of massive stars, as increasingly
strong circumstantial evidence and tantalizing direct evidence suggests
(see, e.g., Lamb 2000), 10-40\% of GRBs may lie at very high redshifts
($z > 5$) (Lamb \& Reichart~2000; see also Ciardi \& Loeb~2000 and
Bromm \& Loeb 2002).

\item
Somewhat surprisingly, the infrared and near-IR afterglows of GRBs are
detectable out to very high redshifts because of cosmological time
dilation (Lamb \& Reichart 2000).

\item
No ``proximity effect'' is expected for GRBs.

\item
GRB afterglows have simple power-law spectra and dramatically outshine
their host galaxies, making it relatively easy to determine the shape
of the red damping wing of the Ly$\alpha$ resonance.

\end{itemize}

\begin{figure}
\plotfiddle{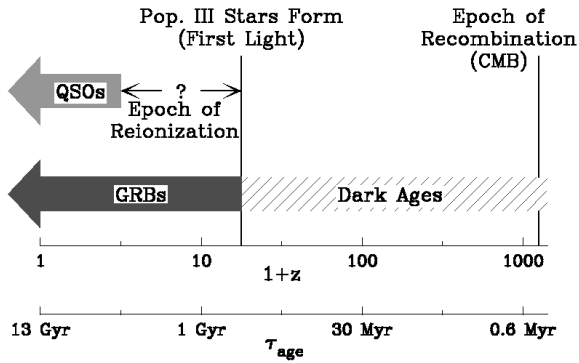}{5.0cm}{0.0}{38.0}{38.0}{-110}{0}
\caption{Cosmological context of very high redshift ($z > 5$) GRBs.
Shown are the epochs of recombination, first light, and re-ionization.
Also shown are the ranges of redshifts corresponding to the ``dark
ages,'' and probed by QSOs and GRBs.  From Lamb (2000).}
\end{figure}

\section{Calculations} 

A stellar population with a Salpeter initial mass function extending
from 0.1 to 120 $M_\odot$ produces $\approx 4000$ ionizing photons per
stellar proton over its lifetime.  
Assuming a (steady) SFR $\dot{M}_*$ and an age $t_*$ for the
star-forming episode, the total number $N_{\rm ion}$ of ionizing
photons produced by the host galaxy is
\begin{equation}
N_{\rm ion} \approx 4.8 \times 10^{70} \left({f_{\rm esc} \over 0.1}
\right) \left({{\dot{M}_*} \over {1000 M_\odot {\rm yr}^{-1}}} \right)
\left({t_8 \over {10^8 {\rm yr}}}\right) \; {\rm ph} \; .
\end{equation}
Here $f_{\rm esc}$ is the fraction of the ionizing photons that are
produced by the stars in the host galaxy and that escape from the
galaxy.  The fraction of ionizing photons that escape from nearby (low
redshift) galaxies is $f_{\rm esc} \sim 0.1$.  However, extinction is
due primarily to dust, which may play a smaller role at redshifts $z
\sim 7$ where the metallicity is expected to be far smaller.  Pop III
stars may produce a factor of $\sim 10$ more ionizing photons, but they
are not expected to be a major contributor to the starlight from
galaxies at redshifts $z = 7-9$.

Consider a GRB host galaxy at redshift $z_{\rm GRB}$ with a (steady)
SFR $\dot{M}_*$ and a star-forming episode of age $t_*$.  We assume the
source is embedded in an initially neutral IGM with a mean density
$\langle \rho_{\rm IGM} \rangle = \Omega_b \rho_{\rm crit} (1+z_{\rm
GRB})^3$.   The radius $R_{S}$ of the ionized photons around a GRB host
galaxy can then be written as $\dot{N}_{\rm ion} t_*$ where $t_*$ is
the age of the star-forming episode.  We have equated the number of
ionizing photons to the number of hydrogen atoms inside $R_S$. This is
valid for low gas clumping factors and small ages, but recombinations
can decrease the size of $R_S$ for $C > 10$ and $t_* > 10^8$ yr.  The
results presented below take the UV spectrum of the GRB afterglow to
have the form $F_\nu = F_0 (\nu/\nu_0))^{-0.5}$.

\section{Results} 

Figure 2 (upper panel) shows the near-IR spectrum of the afterglow of a
GRB in a host galaxy lying at a redshift $z = 6.5$ in which stars have
been forming at a rate $\dot{M}_* = 100 M_\odot$ yr$^{-1}$ for $t_*=
10^8$ yrs.  The nearly horizontal solid line near the top of the panel
shows the adopted intrinsic spectrum, and the bottom solid curve shows
the spectrum including absorption in the IGM and by the neutral atoms
inside the 0.75 Mpc (proper) H II region surrounding the host galaxy of
the GRB.  The lower panel shows optical depth as a function of
wavelength from within the H II region (short-dashed curve), from the
neutral IGM outside the H II region (dotted curve), as well as from the
sum of the two (solid curve).  In both panels, the long-dashed curves
describe an alternative, more realistic treatment of the residual H I
opacity within the H II region (see text).  The arrow indicates the
wavelength of Ly$\alpha$ in the rest frame of the GRB and its host
galaxy.

The shape of the afterglow spectrum of the GRB and the optical depth as
a function of wavelength near the wavelength of Ly$\alpha$ in the rest
frame of the GRB depend sensitively on (1) the clumpiness of the IGM,
(2) the product of the star-formation rate $\dot{M}_\odot$ and the age
$t_8 $of the star-formation episode, and (3) redshift, because of the
rapidly increasing density of the IGM with increasing redshift
[$\rho_{\rm IGM} \propto (1 + z)^3$].  Thus near-IR observations of GRB
afterglows in the vicinity of the wavelength of Ly$\alpha$ in the rest
frame of the GRB can provide unique information

\begin{figure}[t]
\plotfiddle{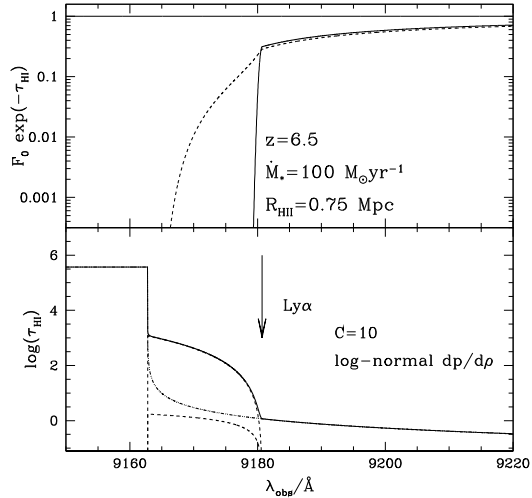}{6.0cm}{0.0}{35.0}{35.0}{-110}{-60}
\caption{Near-IR spectrum of the afterglow of a GRB lying at $z = 6.5$.}
\end{figure}

\section{Conclusions} 

We have shown that NIR and optical observations of GRB afterglows may
provide unique information about the epoch of reionization.  In
particular, high-resolution near-IR observations of GRB afterglows can
not only provide unique information about the properties of the IGM at
very high redshifts, but also about the amount of star-formation that
has occurred prior to the GRB in the host galaxy of the GRB.

\end{document}